\documentclass[%
 reprint,
 amsmath,amssymb,
 aps,
]{revtex4-1}

\usepackage{graphicx}% Include figure files
\usepackage{dcolumn}% Align table columns on decimal point
\usepackage{bm}% bold math
\usepackage{algorithm}
\usepackage{algpseudocode}
\usepackage{todonotes}
\usepackage{placeins}

\newcommand{\ts}{t_s}
\newcommand{\ttt}{t_t}
\newcommand{\xbf}{\mathbf{x}}

\begin{document}

\preprint{APS/123-QED}

\title{Iterative trajectory reweighting for estimation of equilibrium and non-equilibrium observables}

\author{John D. Russo}
\author{Jeremy Copperman}%
 \email{copperma@ohsu.edu}
\author{Daniel M. Zuckerman}
 \email{zuckermd@ohsu.edu}
\affiliation{%
 Department of Biomedical Engineering,
Oregon Health and Science University, Portland, OR
\\
}%

\date{\today}

\begin{abstract}
We present two algorithms by which a set of short, unbiased trajectories can be iteratively reweighted to obtain various observables. 
The first algorithm estimates the stationary (steady state) distribution of a system by iteratively reweighting the trajectories based on the average probability in each state.
The algorithm applies to equilibrium or non-equilibrium steady states, exploiting the `left' stationarity of the distribution under dynamics -- i.e., in a discrete setting, when the column vector of probabilities is multiplied by the transition matrix expressed as a left stochastic matrix.
The second procedure relies on the `right' stationarity of the committor (splitting probability) expressed as a row vector. 
The algorithms are unbiased, do not rely on computing transition matrices, and make no Markov assumption about discretized states. 
Here, we apply the procedures to a one-dimensional double-well potential, and to a 208$\mu$s atomistic Trp-cage folding trajectory from D.E. Shaw Research. 
\end{abstract}

\maketitle

\section{Introduction}
The inability of molecular dynamics (MD) simulation to reach timescales pertinent to complex phenomena in biology and other fields \cite{hollingsworth2018molecular, grossfield2009quantifying, grossfield2018best, zuckerman2011equilibrium, weng2019exploring} has motivated the development of numerous methods to enhance sampling in both equilibrium~\cite{wham, adaptive-biasing, sugita-replex} 
and non-equilibrium~\cite{huber-we, dellago-transition, transition-interface, milestoning, fluxsampling} contexts. 
Markov state models (MSMs) effectively ``stitch together'' shorter trajectories dispersed in configuration space \cite{chodera2014markov,bowman2013introduction} from which both equilibrium and non-equilibrium observables can be computed -- e.g., state populations or kinetic properties.
MSMs can be applied to transition phenomena even when no full, continuous trajectory of a particular transition is present in the original set of trajectories.

This article presents a simple, alternative method for reweighting MSM-like trajectory sets that provides both equilibrium and non-equilibrium information without bias.
The essence of the strategy is to exploit the stationarity of a distribution or property to enable the calculation of that observable in a self-consistent way via iteration.
\emph{The key ingredient is the use of continuous trajectories as the sole basis for analysis, intrinsically accounting for all properties of the underlying dynamics.}
Iteration is employed to reach a fully self-consistent stationary solution.
Trajectory reweighting has previously been applied to biased trajectories (e.g., \cite{zuckerman-dims}) as well as to unbiased trajectories in MSM construction, albeit without self-consistent iteration and under a Markov assumption \cite{voelz-reseeding}.

Observables that can be computed through the iterative approach, without any Markov assumption or lag-time limitation, include the equilibrium distribution, the distribution in a non-equilibrium steady state (NESS), the committor or splitting probability and the mean first-passage time (MFPT) associated with arbitrary macrostates.
The only error in the procedures described below, besides statistical noise, arises from the discretization of phase space into bins.
We emphasize that no Markov assumption is made.

The approach can be understood in the context of estimating the equilibrium distribution based on a set of \emph{unbiased} trajectories initiated from an arbitrary set of initial configurations, presumably out of equilibrium.
For example, the trajectories may be initiated from an approximately uniform distribution in the space of some coordinate of interest.
We assume that a classification of the space has already been performed into bins whose populations are a proxy for the distribution.
Given that the equilibrium distribution does not change in time, if we can assign suitable weights (probabilities) to each of a set of trajectories -- such that the weighted distribution is in equilibrium based on \emph{only} the initial points of each trajectory -- that distribution must remain in equilibrium thereafter.
Although the weights are unknown in advance, they can be set to arbitrary initial values and refined by iteration.

Continuing the equilibrium example, imagine that each trajectory is initially assigned an equal weight, with all weights summing to one.
Now each bin can be monitored over \emph{time}, and the average weight in each bin is recorded.
This average weight is the first non-trivial estimate of the equilibrium probability in the bin.
Physically, bins that attract more trajectories will be assigned larger weights as expected.
In each iteration, the time-averaged probability from the prior iteration is divided among the trajectories which \emph{start} in that bin.
Time-averaged bin probabilities are recomputed and trajectory weights reassigned at each iteration until convergence to steady values.
This procedure is described in Algorithm \ref{algo:steady}.

The same  procedure can be applied for non-equilibrium reweighting.
To obtain the NESS distribution, external and/or boundary conditions must be properly accounted for in preparing trajectories for analysis (Algorithm \ref{algo:traj}), but this is not a significant complication.
With the NESS distribution, the MFPT can be obtained from the Hill relation \cite{hill2005free,bhatt2010steady}.

The committor, also known as the splitting probability \cite{gardiner1985handbook, hill2005free, kampen2007stochastic, weinan-tpt}, exhibits a different type of stationarity \cite{noe} and is estimated by a different but equally simple type of iterative procedure that averages over trajectories instead of bins (Algorithm \ref{algo:comm}).  
Defined as the probability to proceed from a designated initial phase point to a ``target'' macrostate prior to reaching a different ``off-target'' macrostate, the committor can be naively estimated by the fraction of trajectories from the initial point that first reach the target. 
In an iterative approach operating in the space of bins, we can exploit the committor's stationarity: at any fixed time, the average committor of all downstream trajectories emanating from a given bin must match that bin's committor value.
Procedurally, each bin not in the target state is assigned a trivial initial committor estimate of, say, zero. 
A bin's estimate is updated at each iteration as the average over every time step of every trajectory after visiting the bin, with the ``boundary conditions'' that all time points after entering the target macrostate are evaluated as one or, after entering the off-target state, as zero.

We emphasize that these trajectory averaging and reweighting processes make no Markov assumption and are unbiased at the shortest available time discretization.
As with any method, however, the approach is limited by the amount of data which in turn will dictate the sizes of bins which can be used.
More data enables smaller bins and higher phase-space resolution.
Because the dynamics of individual trajectories continually update observable estimates, the discretization error may be less would naively be expected from spatial discretization.

\section{\label{sec:algorithms}Algorithms}

\subsection{Trajectory preparation}

In order to demonstrate our algorithms, we extracted a set of trajectory fragments from one or more long trajectories according to  Algorithm \ref{algo:traj}.
Trajectory fragments may be of fixed length, or variable length if strict absorbing boundary conditions are used.
Source-sink boundary conditions will use spliced fixed-length trajectories.

The analyses performed below are most easily understood based on trajectory fragments sorted by the starting configuration (phase-space point) of each fragment.
These fragments are ``copied'' from the original long trajectory and hence may have overlapping sequences.
For example, fragment 1 may consist of time steps 2 - 101 of the original trajectory, and fragment 2 might be steps 7 - 106.
Correlations are thus introduced, but we estimate statistical uncertainty using fully independent datasets.

\begin{algorithm}[H]
\renewcommand{\thealgorithm}{0}
\caption{Trajectory fragment selection}\label{algo:traj}
\begin{algorithmic}[1]
\State Begin with one or more trajectories, discretized according to a set of bins $i$ (or ``microstates'' in MSM terminology).  For simplicity, we will assume a single long trajectory is used with $t$ denoting the discrete time index.
\State For each bin $i$, generate a list of possible start points $\ts$ which are the time indices of every configuration or phase point within that bin. The set of trajectory starts in bin $i$ -- denoted $\{ t_s \}_i$ -- is not indexed to avoid complex notation.
That is, each of $K_i$ start points indexed by $k= 1 \cdots K_i$ in bin $i$ is fully denoted as $\ts(i,k)$

\If{no absorbing (`open') boundaries} 
    \State The fragments associated with each bin $i$ consist of time points $\ts, \ts+1, \ldots, \ts+M-1$ for each start point in the set $\{ \ts \}_i$.  These fixed-length fragments each have $M$ steps. 
    
\ElsIf{strict absorbing boundary conditions}
    \State Two macrostates consisting of sets of bins should be defined, such that no bin is in more than one macrostate and some bins are ``intermediate'' -- i.e., not in either macrostate.  
    \State The fragments will start \emph{only from intermediate bins} and consist of time points $\ts, \ts+1, \ts+2, \ldots$ for each start point $\ts$.  Each fragment is terminated upon reaching \emph{either} macrostate or at the end of the original trajectory, whichever comes first. 
    
\ElsIf{source-sink boundary conditions}
    \State Two macrostates consisting of sets of bins should be defined, such that no bin is in more than one macrostate and some bins are ``intermediate'' -- i.e., not in either macrostate.  \emph{One macrostate will be the sink (a.k.a. target) and the other is the source state.}
    \State Define a time-independent source distribution $\gamma$ over source bins such that $\sum_j \gamma_j = 1$ with $\gamma_j \geq 0$.
    \State The fragments initially consist of time points $\ts, \ts+1, \ts+2, \ldots \ts+M-1$ for each start point $\ts$.  If the target is reached prior to the final point, let $\ttt$ be the time the target is first reached.
    \State Fragments reaching the target are spliced to fragments starting at the source.  That is, to make a full segment of $M$ steps, the initial list $\ts, \ts+1, \ts+2, \ldots \ttt-1$ is concatenated with a trajectory segment from a source starting point $\ts(j,k)$ with $j \in \mbox{source}$; this segment is re-indexed to start at $\ttt$.  The particular segment is chosen uniformly among the $\ts$ for bin $j$ after $j$ is selected according to $\gamma$.
\EndIf
\end{algorithmic}
\end{algorithm}

\subsection{Equilibrium distribution}

Trajectories can be reweighted into the equilibrium distribution.
Our procedure can be seen as a non-Markovian, fully self-consistent extension of the single-iteration trajectory reweighting recently proposed in a Markov context \cite{voelz-reseeding}.
Reweighting is an old idea \cite{ferrenberg1988new} which is limited by the well-known overlap problem \cite{zuckerman2011equilibrium}.
Overlap remains a concern in any reweighting, but the present strategy uses additional information ignored in many other methods, namely, the dynamical information intrinsic to trajectories.
Algorithm \ref{algo:steady} infers a conformational distribution consistent with the underlying \emph{continuous} dynamics without any Markov assumption.
Discretization necessarily introduces some error but because continuous trajectories evolve irrespective of bin boundaries, this error may be reduced.
That is, trajectory dynamics automatically account for intra-bin landscape features.

Algorithm \ref{algo:steady} uses stationarity of the equilibrium distribution to re-assign weights of trajectory fragments in a self-consistent manner.
In every iteration, the weight of the fragments starting in a given bin is replaced by the time-averaged weight in the bin.
Stationarity is enforced in a self-consistent way because the initial bin probability must match the time average.

\begin{algorithm}[H]
\renewcommand{\thealgorithm}{1}
\caption{Stationary distribution calculation}\label{algo:steady}
\begin{algorithmic}[1]
\State Prepare a set of fixed-length trajectory fragments with open boundary conditions (for equilibrium) or with source-sink conditions (for NESS) following Algorithm \ref{algo:traj}.  Bins not visited by any fragment will be assigned zero probability.  Note that sink/target bins have zero probability by definition.
\State Assign each trajectory fragment an initial weight.  Initial weights are arbitrary, so long as total weight (probability) sums to 1, a condition which is preserved at every time step in every iteration.  Here we assign initial weights so that each bin has equal total initial weight, which is evenly divided among fragments starting in the bin. 
\Repeat
    \ForAll{bins}
        \State {Sum the weights of all fragments in the bin at each time}
        \State The averaged-over-time bin weight is divided equally among trajectory fragments \emph{starting} in that bin for the next iteration.
    \EndFor
\Until{A user-defined convergence threshold is met}
\State The entire iterative procedure can be repeated for trajectory sets generated by progressively trimming the first time-point from each trajectory (to decrease initial state bias), creating a basis for a final estimate averaged over trimmed trajectory sets.  This protocol was not used to generate the data shown.
\State For NESS, the entire iterative procedure can be repeated for trajectory sets generated by progressively trimming the first time-point from each trajectory (to decrease initial state bias), creating a basis for a final estimate averaged over trimmed trajectory sets.  Additionally the source-sink splicing of Algorithm \end{algorithmic}
\end{algorithm}

\subsection{Non-equilibrium steady-state}

The probability distribution of a non-equilibrium steady state (NESS) in the same way (Algorithm \ref{algo:steady}) except that suitable boundary conditions must be enforced.
We focus here on a source-sink NESS because that is most pertinent to rate-constant estimation.
Such a NESS requires defining (i) the absorbing source and sink macrostates, which shall consist strictly of non-overlapping sets of bins and (ii) the source, or feedback, \emph{distribution} $\gamma$ which describes how probability reaching the sink macrostate is redistributed at the source \cite{copperman2019transient}.
In a discrete picture, we let $\gamma_i$ be the fractional probability to be initiated (or fed back) to bin $i$, such that $\sum_i \gamma_i = 1$.
No bin with $\gamma_i >0$ can be part of the sink.
See Algorithm \ref{algo:traj}.

As a technical aside, we note that, somewhat confusingly, bins with positive $\gamma$ values do not in themselves necessarily define the source macrostate.
For example, in the important special case of the source-sink NESS which maintains an equilibrium distribution within the source macrostate (only), bins not on the \emph{surface} of the macrostate strictly require $\gamma=0$ \cite{bhatt2011beyond}.
In any case, our approach applies to arbitrary choices of the source distribution $\gamma$.

\subsection{Committor calculation}

The committor is not a probability distribution per se and exhibits a different kind of stationarity that has been noted previously \cite{noe, weinan-tpt, gardiner1985handbook, kampen2007stochastic}.
The committor $\Pi(\xbf)$ for a phase-space point $\xbf$ is defined to be the probability of trajectories initiated from $\xbf$ reaching a `target' macrostate before reaching a different `initial' macrostate, both of which can be arbitrarily defined if non-overlapping.
We assume dynamics are stochastic and Markovian in the continuous phase space.
Discrete bins used for calculation in the algorithm are not assumed to behave as Markov states.

The iterative algorithm can be understood by first considering `brute force' committor estimation by initiating a large number, $N$, of trajectories from $\xbf$ and computing the fraction which reach the target first.
However, instead of waiting for all such trajectories to be absorbed at one state or the other, we can imagine examining the distribution of phase points $p(\xbf^t)$ at finite time $t$ which evolved from $\xbf$ -- that is, from trajectories initiated at $t=0$ from $\xbf$ with absorbing boundary conditions at initial and target states.
If $t$ is sufficiently short, such that no trajectories have yet been absorbed by either state, the expected fraction that eventually will be absorbed to the target \emph{by definition} is given by the average committor of \emph{current} phase points $\xbf^t$ \cite{noe}.
That is, with trajectories indexed by $i$, the committor can be estimated by
\begin{equation}
    \Pi(\xbf) \doteq (1/N) \sum_i \Pi(\xbf^t_i) \; .
\label{comm}
\end{equation}
This same expression can be used at longer $t$ when some trajectories have been absorbed, if we introduce the `overloaded' definitions $\Pi(\xbf^t_i) \equiv 1$ if trajectory $i$ was absorbed to the target and zero if absorbed to the initial state.
With this adjustment, the estimator \eqref{comm} is applicable at any time $t$.

Algorithm \ref{algo:comm} implements the preceding formulation using an iterative process for self-consistency.
Because the committor average is stationary, we can use \eqref{comm} at any time or by averaging over all times.
Here, committor values are updated based on following trajectories passing through a given phase point, approximated as a discrete bin, and calculating time averages of all the visited `downstream' bins.
By contrast, distribution estimation in Algorithm \ref{algo:steady} averages over time for each bin separately, and do not follow trajectories.
For convenience, trajectories which reach a macrostate are `padded' with committor values of zero or one depending on the macrostate.

Once again, we expect a slight discretization error but using trajectories leverages the maximum possible information about intra-bin dynamics.
Bins are not assumed to exhibit Markovian behavior.

\begin{algorithm}[H]
\renewcommand{\thealgorithm}{2}
\caption{Committor calculation}
\label{algo:comm}
\begin{algorithmic}[1]
\State Begin with a set of absorbing boundary condition trajectory fragments, as described in Algorithm~\ref{algo:traj}
\State Assign each bin within the target macrostate a committor of 1.  All other bins are initialized to 0, including in the initial macrostate.
    \ForAll{trajectory fragments}
        \If{fragment reaches target or initial macrostate}
            \State Pad the trajectory: Assign \emph{fixed} committor values of 1 or 0, respectively, to all time points starting from the absorbing event and ending at the chosen fixed length $M$.
        \EndIf
    \EndFor
\Repeat
    \ForAll{bins}
        \If {bin is within a macrostate}
            \State Do not change committor - it remains 0 or 1
        \Else
        \State The next estimated committor value is the average committor over all bins subsequently visited by all trajectories starting in this bin
        \EndIf
    \EndFor
\Until{Change between iterations is below user-defined convergence threshold}
\end{algorithmic}
\end{algorithm}

\section{\label{sec:results}Systems and Results}

\subsection{\label{sec:results-systems}Systems}
\begin{figure}[b]
    \centering
    \includegraphics[width=0.95\linewidth]{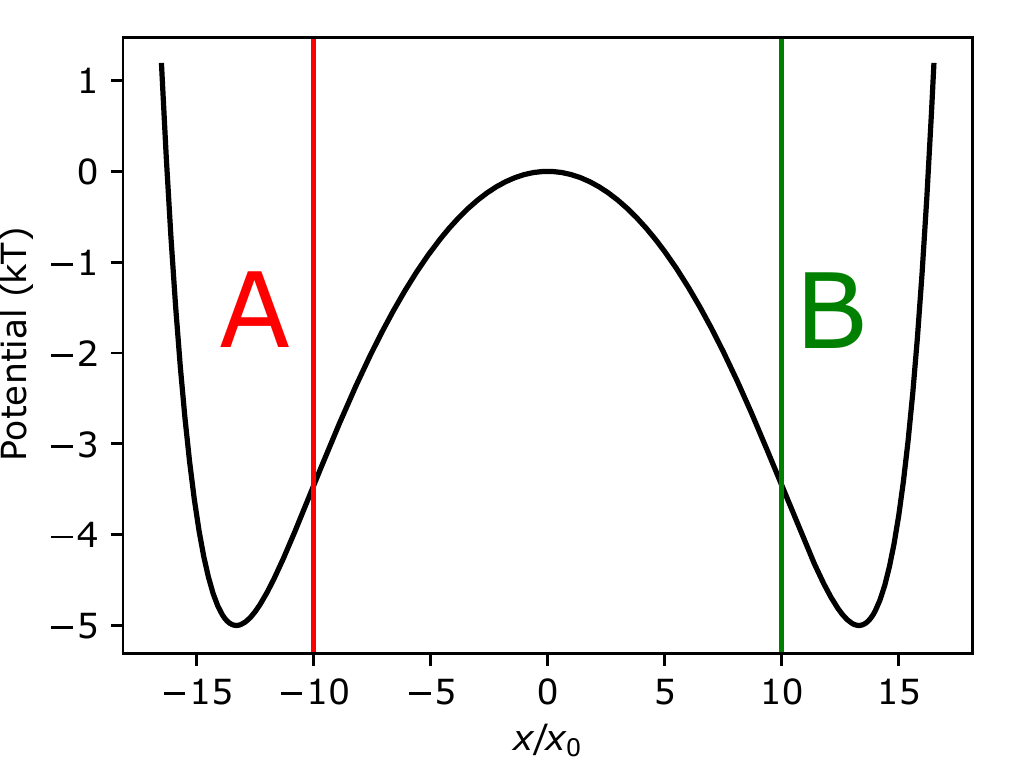}
    \caption{Double-well potential used for overdamped Langevin dynamics simulations. Macrostate A is comprised of states at $x/x_0 < -10$, and B of states at $x/x_0 > 10$. 
    }
    \label{fig:potential}
\end{figure}

The iterative equilibrium distribution estimation technique is first applied to a set of simulated trajectories in 
a one-dimensional (1D) double-well potential with a 5 $k_BT$ barrier, shown in Fig.~\ref{fig:potential} and simulated using overdamped Langevin dynamics.

Motion under overdamped Langevin dynamics obeys 

\begin{equation}
    x_{i+1} = 
    x_i + 
    - \frac{\Delta t }{m \gamma}  \left.\frac{dV}{dx} \right\rvert_{x_i}  
    + \Delta x_{\textrm{rand}}
\end{equation}
where $\gamma = 0.01 \textrm{s}^{-1}$ is the friction coefficient, $m$ is set to 1, $\Delta x_{rand}$ is a stochastic displacement with its magnitude drawn from a Gaussian distribution centered at 0 with $\sigma = \sqrt{2 k_BT \Delta t / m \gamma}$ where $k_BT$ is set to 1 and $\Delta t = 5 \times 10 ^{-4}\textrm{s}$ is the timestep. 
The double-well potential used is given by 
\begin{equation}
V(x) = k_BT \left [ \left(0.1  \frac{x}{x_0} \right)^{10} - \left(0.7 \frac{x}{x_0} \right)^2 \right ].
\label{potl}
\end{equation}
where $x_0$ is an arbitrary reference length.

The full dataset consisted of 32 trajectories, each run for $2 \times 10^6$ steps.
We used 130 equal-width states, of which 80 were in the intermediate region and 25 were in each of states A and B, shown in Fig.~\ref{fig:potential}.

The other system analyzed is a 208 $\mu$s atomistic molecular dynamics simulation of Trp-cage folding saved with 200 ps resolution \cite{lindorff-shawtraj}.
This trajectory is notable for being very long and well-sampled.

\begin{figure}
    \centering
    \includegraphics[width=\linewidth]{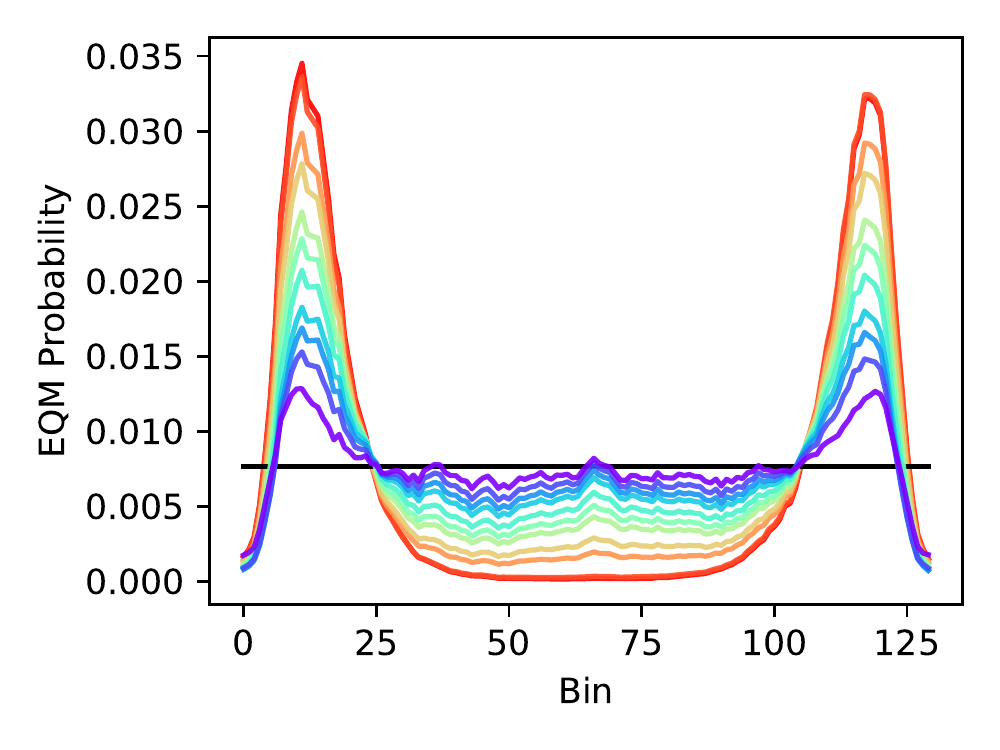}
    \caption{Plot of the iterative equilibrium distribution estimator's convergence. Some intermediate iterations have been omitted for clarity. Warmer colors show later iterations, and the black line is the initial weight in each bin.}
    \label{fig:convergence}
\end{figure}

\subsection{\label{sec:results-eqm}Equilibrium distribution}

\begin{figure}
    \centering
    \includegraphics[width=\linewidth]{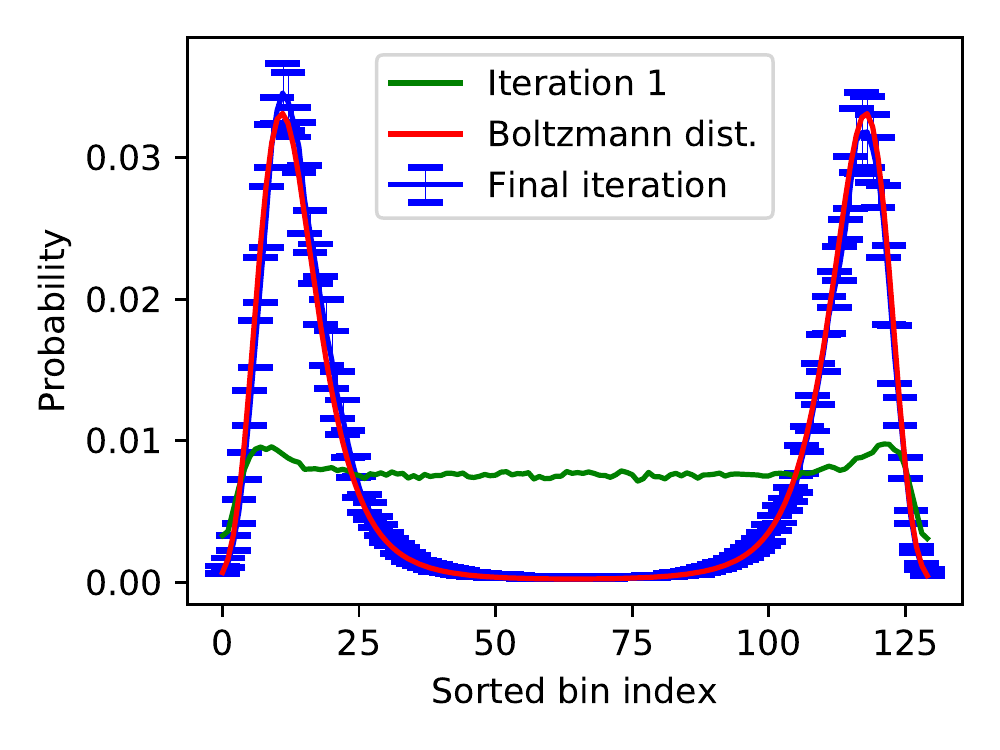}
    \caption{Equilibrium distributions for the double-well potential system. Since the exact form of the potential is known, the Boltzmann distribution (red) provides reference equilibrium probabilities. Shown are the distribution after one iteration (green) and the distribution after the convergence criterion was met (blue). Error bars indicate one standard deviation across 5 independent trials.}
    \label{fig:eqm-doublewell}
\end{figure}

Fig.~\ref{fig:convergence} illustrates the convergence of the iteratively estimated equilibrium distribution and  Fig.~\ref{fig:eqm-doublewell} demonstrates the final result of the iterative  calculation in the 1D double-well system.
In general, the final converged iteration reproduces the Boltzmann distribution precisely and without bias.

\begin{figure}
    \centering
    \includegraphics[width=\linewidth]{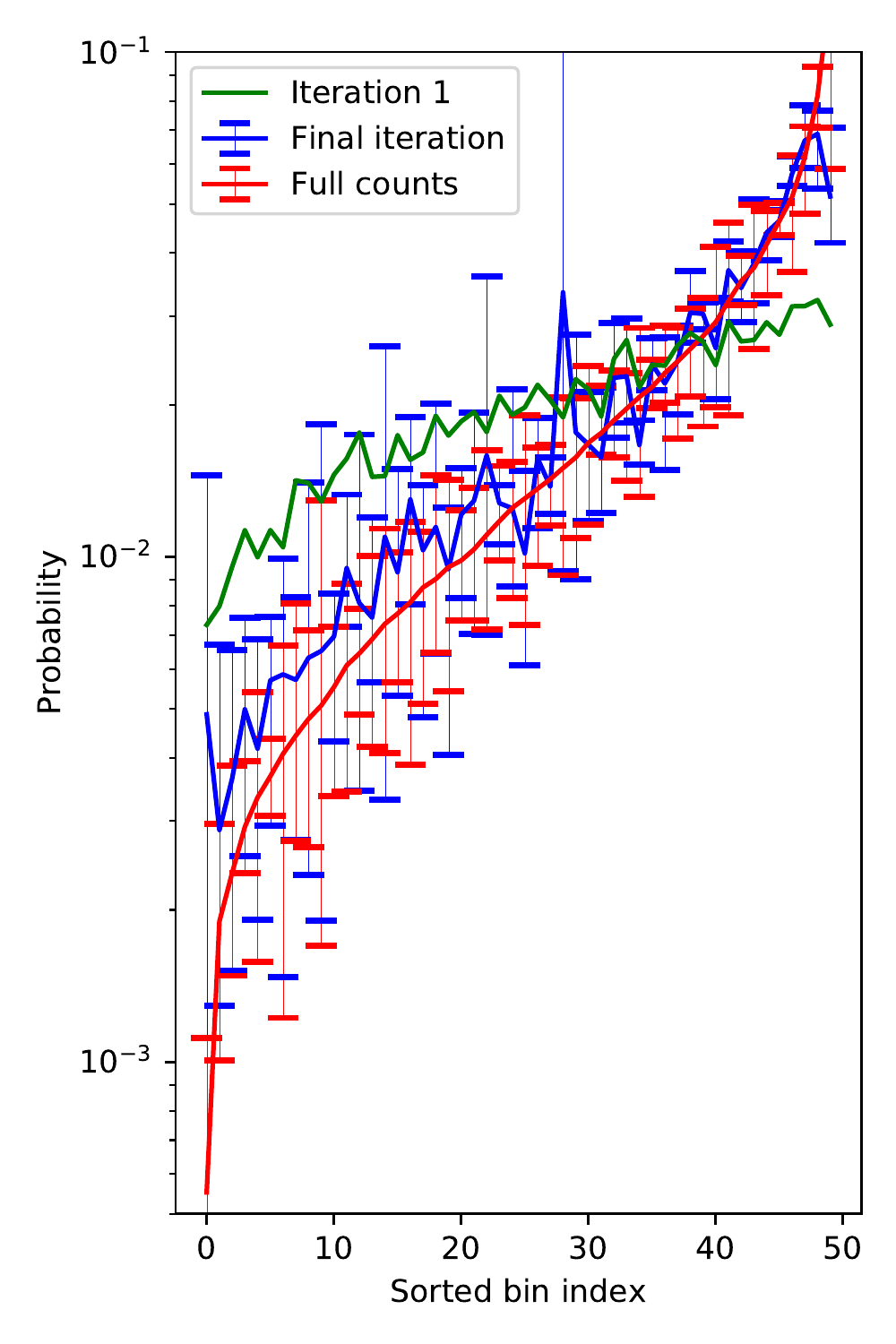}
    \caption{Equilibrium distributions for the Trp-cage folding trajectory fragments, shown on a log and a linear scale. 
    Shown are the distribution after one iteration (green), the distribution after the convergence criterion was met (blue), and counts in each bin from the original full trajectory (red), averaged across independent trials based on sub-dividing the full Shaw trajectory into five segments. 
    Bins have been coarse-grained from 1000 initial bins for visualization. 
    Error bars represent minima and maxima among five independent trials.}
    \label{fig:eqm-shaw}
\end{figure}

Applying the iterative equilibrium distribution estimator to the Trp-cage folding trajectory (Fig.\ \ref{fig:eqm-shaw}) fragments similarly shows reasonable agreement with simple counts. The right-most bin is a notable exception and warrants further investigation.

\FloatBarrier
\subsection{\label{sec:results-comm}Committor calculation}

As before, we first apply the committor estimator to the 1D double-well potential.
With this simple 1D system we are able to directly compute the committor through a ``brute-force"
technique, where a number of trajectories are initialized from each point, and stopped when they
reach a macrostate.
Although the computational cost of this would be prohibitive for a more complex system, 
this is an unbiased reference.

\begin{figure}
    \centering
    \includegraphics[width=\linewidth]{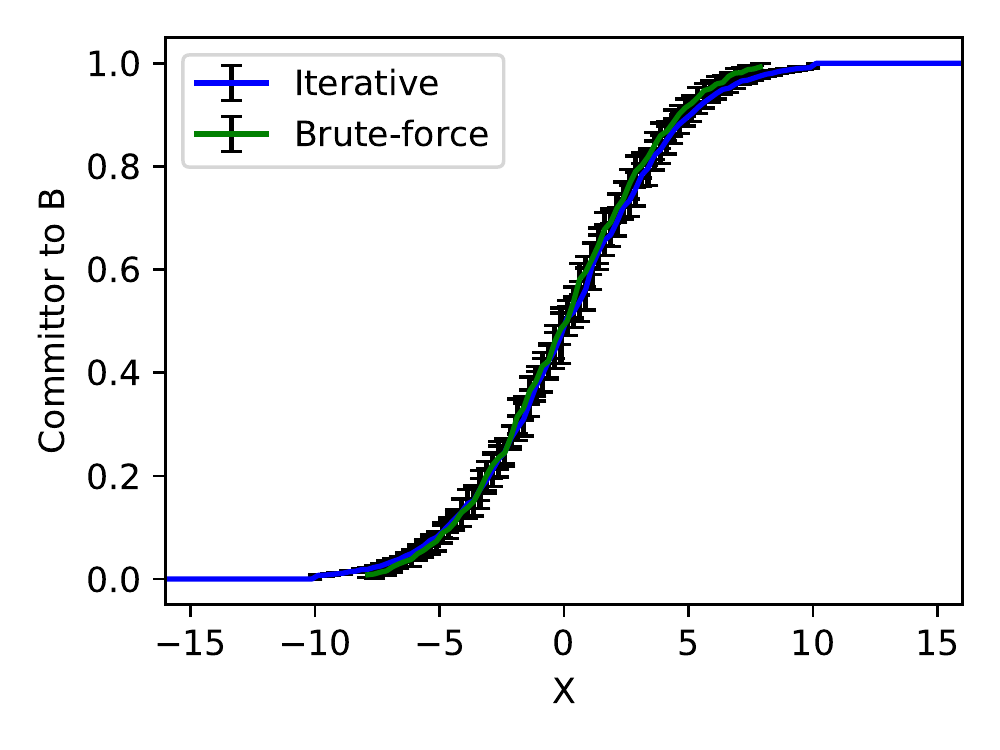}
    \caption{Validation of iterative committor estimation in a one-dimensional model.  Committor estimates are shown for the brute-force/naive calculation (green line) as well as the iterative approach (blue line) vs brute-force result, for the one-dimensional model of the potential in Eq.~\eqref{potl}.  Error bars indicate one standard deviation across 5 independent trials. 
    }
    \label{fig:alg3_doublewell_bf}
\end{figure}

Fig.~\ref{fig:alg3_doublewell_bf} shows the result of the iterative committor estimator along with the 
brute-force reference for the 1D system.
The committor profile follows the expected sigmoid shape between the two wells, with a value of 0.5 at the peak of the barrier.
The iterative approach is thus validated as unbiased, by comparison to brute-force computation.

\begin{figure}[H]
    \centering
    \includegraphics[width=\linewidth]{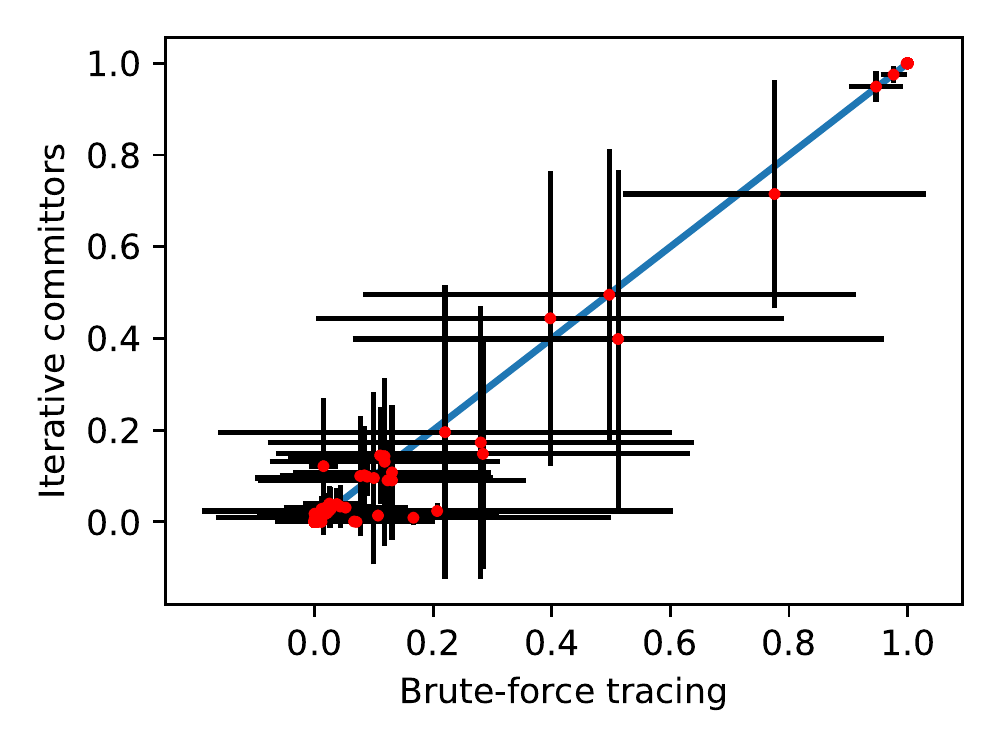}
    \caption{Scatter plot of brute force committor values vs iterative committor values for Trp-cage.
    A line of slope 1 is shown in blue.
    Error bars represent a single standard deviation among independent trials based on sub-dividing the full Shaw trajectory into five segments.
    }
    \label{fig:alg3_scatter}
\end{figure}

We also applied the iterative scheme to estimating the committor in for the Trp-cage system.
Once again, brute-force reference committor values were obtained by following trajectory fragments originating in each bin until they reached a macrostate; the fraction that reaching state B before state A determined the committor.
As seen in Fig.~\ref{fig:alg3_scatter}, the iterative committor estimation algorithm yields results for the Trp-cage data that track these brute-force estimates well, especially near the macrostates.

\section{Conclusions}
We have introduced algorithms that employ two well-known principles, iteration and stationarity, to estimate key observables from a trajectory or set of trajectories.
In principle, the input trajectories need not follow any prescribed distribution.
The procedures described do not rely on a Markov assumption.
Although discrete bins are used for ``accounting,'' the continuous trajectories embody all details of the landscape and dynamics which, in turn, are included implicitly in the analyses.

Subsequent work will show that the procedures described here are formally equivalent to 'power method' \cite{powerMethod} evaluation of the stationary distribution of a non-standard transition matrix that accounts for trajectory dynamics over all available timescales, as pointed out to us by David Aristoff and Gideon Simpson.

\begin{acknowledgments}
We appreciate helpful discussions with David Aristoff and Gideon Simpson.
We thank DE Shaw Research for sharing the protein folding trajectory with us and the NIH for support through Grant GM115805.
\end{acknowledgments}

\FloatBarrier

\bibliography{bibliography}

\providecommand{\noopsort}[1]{}\providecommand{\singleletter}[1]{#1}%
\begin{thebibliography}{28}%
\makeatletter
\providecommand \@ifxundefined [1]{%
 \@ifx{#1\undefined}
}%
\providecommand \@ifnum [1]{%
 \ifnum #1\expandafter \@firstoftwo
 \else \expandafter \@secondoftwo
 \fi
}%
\providecommand \@ifx [1]{%
 \ifx #1\expandafter \@firstoftwo
 \else \expandafter \@secondoftwo
 \fi
}%
\providecommand \natexlab [1]{#1}%
\providecommand \enquote  [1]{``#1''}%
\providecommand \bibnamefont  [1]{#1}%
\providecommand \bibfnamefont [1]{#1}%
\providecommand \citenamefont [1]{#1}%
\providecommand \href@noop [0]{\@secondoftwo}%
\providecommand \href [0]{\begingroup \@sanitize@url \@href}%
\providecommand \@href[1]{\@@startlink{#1}\@@href}%
\providecommand \@@href[1]{\endgroup#1\@@endlink}%
\providecommand \@sanitize@url [0]{\catcode `\\12\catcode `\$12\catcode
  `\&12\catcode `\#12\catcode `\^12\catcode `\_12\catcode `\%12\relax}%
\providecommand \@@startlink[1]{}%
\providecommand \@@endlink[0]{}%
\providecommand \url  [0]{\begingroup\@sanitize@url \@url }%
\providecommand \@url [1]{\endgroup\@href {#1}{\urlprefix }}%
\providecommand \urlprefix  [0]{URL }%
\providecommand \Eprint [0]{\href }%
\providecommand \doibase [0]{https://doi.org/}%
\providecommand \selectlanguage [0]{\@gobble}%
\providecommand \bibinfo  [0]{\@secondoftwo}%
\providecommand \bibfield  [0]{\@secondoftwo}%
\providecommand \translation [1]{[#1]}%
\providecommand \BibitemOpen [0]{}%
\providecommand \bibitemStop [0]{}%
\providecommand \bibitemNoStop [0]{.\EOS\space}%
\providecommand \EOS [0]{\spacefactor3000\relax}%
\providecommand \BibitemShut  [1]{\csname bibitem#1\endcsname}%
\let\auto@bib@innerbib\@empty
%</preamble>
\bibitem [{\citenamefont {Hollingsworth}\ and\ \citenamefont
  {Dror}(2018)}]{hollingsworth2018molecular}%
  \BibitemOpen
  \bibfield  {author} {\bibinfo {author} {\bibfnamefont {S.~A.}\ \bibnamefont
  {Hollingsworth}}\ and\ \bibinfo {author} {\bibfnamefont {R.~O.}\ \bibnamefont
  {Dror}},\ }\bibfield  {title} {\bibinfo {title} {Molecular dynamics
  simulation for all},\ }\href@noop {} {\bibfield  {journal} {\bibinfo
  {journal} {Neuron}\ }\textbf {\bibinfo {volume} {99}},\ \bibinfo {pages}
  {1129} (\bibinfo {year} {2018})}\BibitemShut {NoStop}%
\bibitem [{\citenamefont {Grossfield}\ and\ \citenamefont
  {Zuckerman}(2009)}]{grossfield2009quantifying}%
  \BibitemOpen
  \bibfield  {author} {\bibinfo {author} {\bibfnamefont {A.}~\bibnamefont
  {Grossfield}}\ and\ \bibinfo {author} {\bibfnamefont {D.~M.}\ \bibnamefont
  {Zuckerman}},\ }\bibfield  {title} {\bibinfo {title} {Quantifying uncertainty
  and sampling quality in biomolecular simulations},\ }\href@noop {} {\bibfield
   {journal} {\bibinfo  {journal} {Annual reports in computational chemistry}\
  }\textbf {\bibinfo {volume} {5}},\ \bibinfo {pages} {23} (\bibinfo {year}
  {2009})}\BibitemShut {NoStop}%
\bibitem [{\citenamefont {Grossfield}\ \emph {et~al.}(2018)\citenamefont
  {Grossfield}, \citenamefont {Patrone}, \citenamefont {Roe}, \citenamefont
  {Schultz}, \citenamefont {Siderius},\ and\ \citenamefont
  {Zuckerman}}]{grossfield2018best}%
  \BibitemOpen
  \bibfield  {author} {\bibinfo {author} {\bibfnamefont {A.}~\bibnamefont
  {Grossfield}}, \bibinfo {author} {\bibfnamefont {P.~N.}\ \bibnamefont
  {Patrone}}, \bibinfo {author} {\bibfnamefont {D.~R.}\ \bibnamefont {Roe}},
  \bibinfo {author} {\bibfnamefont {A.~J.}\ \bibnamefont {Schultz}}, \bibinfo
  {author} {\bibfnamefont {D.~W.}\ \bibnamefont {Siderius}},\ and\ \bibinfo
  {author} {\bibfnamefont {D.~M.}\ \bibnamefont {Zuckerman}},\ }\bibfield
  {title} {\bibinfo {title} {Best practices for quantification of uncertainty
  and sampling quality in molecular simulations [article v1. 0]},\ }\href@noop
  {} {\bibfield  {journal} {\bibinfo  {journal} {Living journal of
  computational molecular science}\ }\textbf {\bibinfo {volume} {1}} (\bibinfo
  {year} {2018})}\BibitemShut {NoStop}%
\bibitem [{\citenamefont {Zuckerman}(2011)}]{zuckerman2011equilibrium}%
  \BibitemOpen
  \bibfield  {author} {\bibinfo {author} {\bibfnamefont {D.~M.}\ \bibnamefont
  {Zuckerman}},\ }\bibfield  {title} {\bibinfo {title} {Equilibrium sampling in
  biomolecular simulations},\ }\href@noop {} {\bibfield  {journal} {\bibinfo
  {journal} {Annual review of biophysics}\ }\textbf {\bibinfo {volume} {40}},\
  \bibinfo {pages} {41} (\bibinfo {year} {2011})}\BibitemShut {NoStop}%
\bibitem [{\citenamefont {Weng}\ \emph {et~al.}(2019)\citenamefont {Weng},
  \citenamefont {Stott},\ and\ \citenamefont {Toner}}]{weng2019exploring}%
  \BibitemOpen
  \bibfield  {author} {\bibinfo {author} {\bibfnamefont {L.}~\bibnamefont
  {Weng}}, \bibinfo {author} {\bibfnamefont {S.~L.}\ \bibnamefont {Stott}},\
  and\ \bibinfo {author} {\bibfnamefont {M.}~\bibnamefont {Toner}},\ }\bibfield
   {title} {\bibinfo {title} {Exploring dynamics and structure of biomolecules,
  cryoprotectants, and water using molecular dynamics simulations: implications
  for biostabilization and biopreservation},\ }\href@noop {} {\bibfield
  {journal} {\bibinfo  {journal} {Annual review of biomedical engineering}\
  }\textbf {\bibinfo {volume} {21}},\ \bibinfo {pages} {1} (\bibinfo {year}
  {2019})}\BibitemShut {NoStop}%
\bibitem [{\citenamefont {Kumar}\ \emph {et~al.}(1992)\citenamefont {Kumar},
  \citenamefont {Rosenberg}, \citenamefont {Bouzida}, \citenamefont
  {Swendsen},\ and\ \citenamefont {Kollman}}]{wham}%
  \BibitemOpen
  \bibfield  {author} {\bibinfo {author} {\bibfnamefont {S.}~\bibnamefont
  {Kumar}}, \bibinfo {author} {\bibfnamefont {J.~M.}\ \bibnamefont
  {Rosenberg}}, \bibinfo {author} {\bibfnamefont {D.}~\bibnamefont {Bouzida}},
  \bibinfo {author} {\bibfnamefont {R.~H.}\ \bibnamefont {Swendsen}},\ and\
  \bibinfo {author} {\bibfnamefont {P.~A.}\ \bibnamefont {Kollman}},\
  }\bibfield  {title} {\bibinfo {title} {{THE weighted histogram analysis
  method for free-energy calculations on biomolecules. I. The method}},\ }\href
  {https://doi.org/10.1002/jcc.540130812} {\bibfield  {journal} {\bibinfo
  {journal} {Journal of Computational Chemistry}\ }\textbf {\bibinfo {volume}
  {13}},\ \bibinfo {pages} {1011} (\bibinfo {year} {1992})}\BibitemShut
  {NoStop}%
\bibitem [{\citenamefont {Darve}\ \emph {et~al.}(2008)\citenamefont {Darve},
  \citenamefont {Rodr{\'{i}}guez-G{\'{o}}mez},\ and\ \citenamefont
  {Pohorille}}]{adaptive-biasing}%
  \BibitemOpen
  \bibfield  {author} {\bibinfo {author} {\bibfnamefont {E.}~\bibnamefont
  {Darve}}, \bibinfo {author} {\bibfnamefont {D.}~\bibnamefont
  {Rodr{\'{i}}guez-G{\'{o}}mez}},\ and\ \bibinfo {author} {\bibfnamefont
  {A.}~\bibnamefont {Pohorille}},\ }\bibfield  {title} {\bibinfo {title}
  {{Adaptive biasing force method for scalar and vector free energy
  calculations}},\ }\href {https://doi.org/10.1063/1.2829861} {\bibfield
  {journal} {\bibinfo  {journal} {The Journal of Chemical Physics}\ }\textbf
  {\bibinfo {volume} {128}},\ \bibinfo {pages} {144120} (\bibinfo {year}
  {2008})}\BibitemShut {NoStop}%
\bibitem [{\citenamefont {Sugita}\ and\ \citenamefont
  {Okamoto}(1999)}]{sugita-replex}%
  \BibitemOpen
  \bibfield  {author} {\bibinfo {author} {\bibfnamefont {Y.}~\bibnamefont
  {Sugita}}\ and\ \bibinfo {author} {\bibfnamefont {Y.}~\bibnamefont
  {Okamoto}},\ }\bibfield  {title} {\bibinfo {title} {{Replica-exchange
  molecular dynamics method for protein folding}},\ }\href
  {https://doi.org/https://doi.org/10.1016/S0009-2614(99)01123-9} {\bibfield
  {journal} {\bibinfo  {journal} {Chemical Physics Letters}\ }\textbf {\bibinfo
  {volume} {314}},\ \bibinfo {pages} {141} (\bibinfo {year}
  {1999})}\BibitemShut {NoStop}%
\bibitem [{\citenamefont {Huber}\ and\ \citenamefont {Kim}(1996)}]{huber-we}%
  \BibitemOpen
  \bibfield  {author} {\bibinfo {author} {\bibfnamefont {G.~A.}\ \bibnamefont
  {Huber}}\ and\ \bibinfo {author} {\bibfnamefont {S.}~\bibnamefont {Kim}},\
  }\bibfield  {title} {\bibinfo {title} {{Weighted-ensemble Brownian dynamics
  simulations for protein association reactions}},\ }\href
  {https://doi.org/https://doi.org/10.1016/S0006-3495(96)79552-8} {\bibfield
  {journal} {\bibinfo  {journal} {Biophysical Journal}\ }\textbf {\bibinfo
  {volume} {70}},\ \bibinfo {pages} {97} (\bibinfo {year} {1996})}\BibitemShut
  {NoStop}%
\bibitem [{\citenamefont {Dellago}\ \emph {et~al.}(2002)\citenamefont
  {Dellago}, \citenamefont {Bolhuis},\ and\ \citenamefont
  {Geissler}}]{dellago-transition}%
  \BibitemOpen
  \bibfield  {author} {\bibinfo {author} {\bibfnamefont {C.}~\bibnamefont
  {Dellago}}, \bibinfo {author} {\bibfnamefont {P.}~\bibnamefont {Bolhuis}},\
  and\ \bibinfo {author} {\bibfnamefont {P.~L.}\ \bibnamefont {Geissler}},\
  }\bibfield  {title} {\bibinfo {title} {Transition path sampling},\
  }\href@noop {} {\bibfield  {journal} {\bibinfo  {journal} {Advances in
  chemical physics}\ }\textbf {\bibinfo {volume} {123}},\ \bibinfo {pages} {1}
  (\bibinfo {year} {2002})}\BibitemShut {NoStop}%
\bibitem [{\citenamefont {van Erp}\ \emph {et~al.}(2003)\citenamefont {van
  Erp}, \citenamefont {Moroni},\ and\ \citenamefont
  {Bolhuis}}]{transition-interface}%
  \BibitemOpen
  \bibfield  {author} {\bibinfo {author} {\bibfnamefont {T.~S.}\ \bibnamefont
  {van Erp}}, \bibinfo {author} {\bibfnamefont {D.}~\bibnamefont {Moroni}},\
  and\ \bibinfo {author} {\bibfnamefont {P.~G.}\ \bibnamefont {Bolhuis}},\
  }\bibfield  {title} {\bibinfo {title} {{A novel path sampling method for the
  calculation of rate constants}},\ }\href {https://doi.org/10.1063/1.1562614}
  {\bibfield  {journal} {\bibinfo  {journal} {The Journal of Chemical Physics}\
  }\textbf {\bibinfo {volume} {118}},\ \bibinfo {pages} {7762} (\bibinfo {year}
  {2003})}\BibitemShut {NoStop}%
\bibitem [{\citenamefont {Faradjian}\ and\ \citenamefont
  {Elber}(2004)}]{milestoning}%
  \BibitemOpen
  \bibfield  {author} {\bibinfo {author} {\bibfnamefont {A.~K.}\ \bibnamefont
  {Faradjian}}\ and\ \bibinfo {author} {\bibfnamefont {R.}~\bibnamefont
  {Elber}},\ }\bibfield  {title} {\bibinfo {title} {{Computing time scales from
  reaction coordinates by milestoning}},\ }\href
  {https://doi.org/10.1063/1.1738640} {\bibfield  {journal} {\bibinfo
  {journal} {The Journal of Chemical Physics}\ }\textbf {\bibinfo {volume}
  {120}},\ \bibinfo {pages} {10880} (\bibinfo {year} {2004})}\BibitemShut
  {NoStop}%
\bibitem [{\citenamefont {Allen}\ \emph {et~al.}(2006)\citenamefont {Allen},
  \citenamefont {Frenkel},\ and\ \citenamefont {ten Wolde}}]{fluxsampling}%
  \BibitemOpen
  \bibfield  {author} {\bibinfo {author} {\bibfnamefont {R.~J.}\ \bibnamefont
  {Allen}}, \bibinfo {author} {\bibfnamefont {D.}~\bibnamefont {Frenkel}},\
  and\ \bibinfo {author} {\bibfnamefont {P.~R.}\ \bibnamefont {ten Wolde}},\
  }\bibfield  {title} {\bibinfo {title} {{Simulating rare events in equilibrium
  or nonequilibrium stochastic systems}},\ }\href
  {https://doi.org/10.1063/1.2140273} {\bibfield  {journal} {\bibinfo
  {journal} {The Journal of Chemical Physics}\ }\textbf {\bibinfo {volume}
  {124}},\ \bibinfo {pages} {24102} (\bibinfo {year} {2006})}\BibitemShut
  {NoStop}%
\bibitem [{\citenamefont {Chodera}\ and\ \citenamefont
  {No{\'e}}(2014)}]{chodera2014markov}%
  \BibitemOpen
  \bibfield  {author} {\bibinfo {author} {\bibfnamefont {J.~D.}\ \bibnamefont
  {Chodera}}\ and\ \bibinfo {author} {\bibfnamefont {F.}~\bibnamefont
  {No{\'e}}},\ }\bibfield  {title} {\bibinfo {title} {Markov state models of
  biomolecular conformational dynamics},\ }\href@noop {} {\bibfield  {journal}
  {\bibinfo  {journal} {Current opinion in structural biology}\ }\textbf
  {\bibinfo {volume} {25}},\ \bibinfo {pages} {135} (\bibinfo {year}
  {2014})}\BibitemShut {NoStop}%
\bibitem [{\citenamefont {Bowman}\ \emph {et~al.}(2013)\citenamefont {Bowman},
  \citenamefont {Pande},\ and\ \citenamefont
  {No{\'e}}}]{bowman2013introduction}%
  \BibitemOpen
  \bibfield  {author} {\bibinfo {author} {\bibfnamefont {G.~R.}\ \bibnamefont
  {Bowman}}, \bibinfo {author} {\bibfnamefont {V.~S.}\ \bibnamefont {Pande}},\
  and\ \bibinfo {author} {\bibfnamefont {F.}~\bibnamefont {No{\'e}}},\
  }\href@noop {} {\emph {\bibinfo {title} {An introduction to Markov state
  models and their application to long timescale molecular simulation}}},\
  Vol.\ \bibinfo {volume} {797}\ (\bibinfo  {publisher} {Springer Science \&
  Business Media},\ \bibinfo {year} {2013})\BibitemShut {NoStop}%
\bibitem [{\citenamefont {Zuckerman}\ and\ \citenamefont
  {Woolf}(2000)}]{zuckerman-dims}%
  \BibitemOpen
  \bibfield  {author} {\bibinfo {author} {\bibfnamefont {D.~M.}\ \bibnamefont
  {Zuckerman}}\ and\ \bibinfo {author} {\bibfnamefont {T.~B.}\ \bibnamefont
  {Woolf}},\ }\bibfield  {title} {\bibinfo {title} {Efficient dynamic
  importance sampling of rare events in one dimension},\ }\href
  {https://doi.org/10.1103/PhysRevE.63.016702} {\bibfield  {journal} {\bibinfo
  {journal} {Phys. Rev. E}\ }\textbf {\bibinfo {volume} {63}},\ \bibinfo
  {pages} {016702} (\bibinfo {year} {2000})}\BibitemShut {NoStop}%
\bibitem [{\citenamefont {Wan}\ and\ \citenamefont
  {Voelz}(2020)}]{voelz-reseeding}%
  \BibitemOpen
  \bibfield  {author} {\bibinfo {author} {\bibfnamefont {H.}~\bibnamefont
  {Wan}}\ and\ \bibinfo {author} {\bibfnamefont {V.~A.}\ \bibnamefont
  {Voelz}},\ }\bibfield  {title} {\bibinfo {title} {{Adaptive Markov state
  model estimation using short reseeding trajectories}},\ }\href
  {https://doi.org/10.1063/1.5142457} {\bibfield  {journal} {\bibinfo
  {journal} {The Journal of Chemical Physics}\ }\textbf {\bibinfo {volume}
  {152}},\ \bibinfo {pages} {24103} (\bibinfo {year} {2020})}\BibitemShut
  {NoStop}%
\bibitem [{\citenamefont {Hill}(2005)}]{hill2005free}%
  \BibitemOpen
  \bibfield  {author} {\bibinfo {author} {\bibfnamefont {T.}~\bibnamefont
  {Hill}},\ }\href {https://books.google.com/books?id=AqVsAAAACAAJ} {\emph
  {\bibinfo {title} {Free Energy Transduction and Biochemical Cycle
  Kinetics}}},\ Dover Books on Chemistry\ (\bibinfo  {publisher} {Dover
  Publications},\ \bibinfo {year} {2005})\BibitemShut {NoStop}%
\bibitem [{\citenamefont {Bhatt}\ \emph {et~al.}(2010)\citenamefont {Bhatt},
  \citenamefont {Zhang},\ and\ \citenamefont {Zuckerman}}]{bhatt2010steady}%
  \BibitemOpen
  \bibfield  {author} {\bibinfo {author} {\bibfnamefont {D.}~\bibnamefont
  {Bhatt}}, \bibinfo {author} {\bibfnamefont {B.~W.}\ \bibnamefont {Zhang}},\
  and\ \bibinfo {author} {\bibfnamefont {D.~M.}\ \bibnamefont {Zuckerman}},\
  }\bibfield  {title} {\bibinfo {title} {Steady-state simulations using
  weighted ensemble path sampling},\ }\href@noop {} {\bibfield  {journal}
  {\bibinfo  {journal} {The Journal of chemical physics}\ }\textbf {\bibinfo
  {volume} {133}},\ \bibinfo {pages} {014110} (\bibinfo {year}
  {2010})}\BibitemShut {NoStop}%
\bibitem [{\citenamefont {Gardiner}(1985)}]{gardiner1985handbook}%
  \BibitemOpen
  \bibfield  {author} {\bibinfo {author} {\bibfnamefont {C.~W.}\ \bibnamefont
  {Gardiner}},\ }\href@noop {} {\emph {\bibinfo {title} {Handbook of stochastic
  methods for physics, chemistry, and the natural sciences}}}\ (\bibinfo
  {publisher} {Springer-Verlag},\ \bibinfo {address} {Berlin New York},\
  \bibinfo {year} {1985})\BibitemShut {NoStop}%
\bibitem [{\citenamefont {Van~Kampen}(2007)}]{kampen2007stochastic}%
  \BibitemOpen
  \bibfield  {author} {\bibinfo {author} {\bibfnamefont {N.~G.}\ \bibnamefont
  {Van~Kampen}},\ }\href@noop {} {\emph {\bibinfo {title} {Stochastic processes
  in physics and chemistry}}}\ (\bibinfo  {publisher} {Elsevier},\ \bibinfo
  {address} {Amsterdam Boston London},\ \bibinfo {year} {2007})\BibitemShut
  {NoStop}%
\bibitem [{\citenamefont {E.}\ and\ \citenamefont
  {Vanden-Eijnden}(2006)}]{weinan-tpt}%
  \BibitemOpen
  \bibfield  {author} {\bibinfo {author} {\bibfnamefont {W.}~\bibnamefont
  {E.}}\ and\ \bibinfo {author} {\bibfnamefont {E.}~\bibnamefont
  {Vanden-Eijnden}},\ }\bibfield  {title} {\bibinfo {title} {{Towards a Theory
  of Transition Paths}},\ }\href {https://doi.org/10.1007/s10955-005-9003-9}
  {\bibfield  {journal} {\bibinfo  {journal} {Journal of Statistical Physics}\
  }\textbf {\bibinfo {volume} {123}},\ \bibinfo {pages} {503} (\bibinfo {year}
  {2006})}\BibitemShut {NoStop}%
\bibitem [{\citenamefont {Prinz}\ \emph {et~al.}(2011)\citenamefont {Prinz},
  \citenamefont {Held}, \citenamefont {Smith},\ and\ \citenamefont
  {Noé}}]{noe}%
  \BibitemOpen
  \bibfield  {author} {\bibinfo {author} {\bibfnamefont {J.-H.}\ \bibnamefont
  {Prinz}}, \bibinfo {author} {\bibfnamefont {M.}~\bibnamefont {Held}},
  \bibinfo {author} {\bibfnamefont {J.~C.}\ \bibnamefont {Smith}},\ and\
  \bibinfo {author} {\bibfnamefont {F.}~\bibnamefont {Noé}},\ }\bibfield
  {title} {\bibinfo {title} {Efficient computation, sensitivity, and error
  analysis of committor probabilities for complex dynamical processes},\ }\href
  {https://doi.org/10.1137/100789191} {\bibfield  {journal} {\bibinfo
  {journal} {Multiscale Modeling \& Simulation}\ }\textbf {\bibinfo {volume}
  {9}},\ \bibinfo {pages} {545} (\bibinfo {year} {2011})},\ \Eprint
  {https://arxiv.org/abs/https://doi.org/10.1137/100789191}
  {https://doi.org/10.1137/100789191} \BibitemShut {NoStop}%
\bibitem [{\citenamefont {Ferrenberg}\ and\ \citenamefont
  {Swendsen}(1988)}]{ferrenberg1988new}%
  \BibitemOpen
  \bibfield  {author} {\bibinfo {author} {\bibfnamefont {A.~M.}\ \bibnamefont
  {Ferrenberg}}\ and\ \bibinfo {author} {\bibfnamefont {R.~H.}\ \bibnamefont
  {Swendsen}},\ }\bibfield  {title} {\bibinfo {title} {New monte carlo
  technique for studying phase transitions},\ }\href@noop {} {\bibfield
  {journal} {\bibinfo  {journal} {Physical review letters}\ }\textbf {\bibinfo
  {volume} {61}},\ \bibinfo {pages} {2635} (\bibinfo {year}
  {1988})}\BibitemShut {NoStop}%
\bibitem [{\citenamefont {Copperman}\ \emph {et~al.}(2019)\citenamefont
  {Copperman}, \citenamefont {Aristoff}, \citenamefont {Makarov}, \citenamefont
  {Simpson},\ and\ \citenamefont {Zuckerman}}]{copperman2019transient}%
  \BibitemOpen
  \bibfield  {author} {\bibinfo {author} {\bibfnamefont {J.}~\bibnamefont
  {Copperman}}, \bibinfo {author} {\bibfnamefont {D.}~\bibnamefont {Aristoff}},
  \bibinfo {author} {\bibfnamefont {D.~E.}\ \bibnamefont {Makarov}}, \bibinfo
  {author} {\bibfnamefont {G.}~\bibnamefont {Simpson}},\ and\ \bibinfo {author}
  {\bibfnamefont {D.~M.}\ \bibnamefont {Zuckerman}},\ }\bibfield  {title}
  {\bibinfo {title} {Transient probability currents provide upper and lower
  bounds on non-equilibrium steady-state currents in the smoluchowski
  picture},\ }\href@noop {} {\bibfield  {journal} {\bibinfo  {journal} {The
  Journal of chemical physics}\ }\textbf {\bibinfo {volume} {151}},\ \bibinfo
  {pages} {174108} (\bibinfo {year} {2019})}\BibitemShut {NoStop}%
\bibitem [{\citenamefont {Bhatt}\ and\ \citenamefont
  {Zuckerman}(2011)}]{bhatt2011beyond}%
  \BibitemOpen
  \bibfield  {author} {\bibinfo {author} {\bibfnamefont {D.}~\bibnamefont
  {Bhatt}}\ and\ \bibinfo {author} {\bibfnamefont {D.~M.}\ \bibnamefont
  {Zuckerman}},\ }\bibfield  {title} {\bibinfo {title} {Beyond microscopic
  reversibility: Are observable nonequilibrium processes precisely
  reversible?},\ }\href@noop {} {\bibfield  {journal} {\bibinfo  {journal}
  {Journal of chemical theory and computation}\ }\textbf {\bibinfo {volume}
  {7}},\ \bibinfo {pages} {2520} (\bibinfo {year} {2011})}\BibitemShut
  {NoStop}%
\bibitem [{\citenamefont {Lindorff-Larsen}\ \emph {et~al.}(2011)\citenamefont
  {Lindorff-Larsen}, \citenamefont {Piana}, \citenamefont {Dror},\ and\
  \citenamefont {Shaw}}]{lindorff-shawtraj}%
  \BibitemOpen
  \bibfield  {author} {\bibinfo {author} {\bibfnamefont {K.}~\bibnamefont
  {Lindorff-Larsen}}, \bibinfo {author} {\bibfnamefont {S.}~\bibnamefont
  {Piana}}, \bibinfo {author} {\bibfnamefont {R.~O.}\ \bibnamefont {Dror}},\
  and\ \bibinfo {author} {\bibfnamefont {D.~E.}\ \bibnamefont {Shaw}},\
  }\bibfield  {title} {\bibinfo {title} {{How Fast-Folding Proteins Fold}},\
  }\href {https://doi.org/10.1126/science.1208351} {\bibfield  {journal}
  {\bibinfo  {journal} {Science}\ }\textbf {\bibinfo {volume} {334}},\ \bibinfo
  {pages} {517} (\bibinfo {year} {2011})}\BibitemShut {NoStop}%
\bibitem [{pow()}]{powerMethod}%
  \BibitemOpen
  \href@noop {} {\bibinfo {title} {Power iteration}},\ \bibinfo {howpublished}
  {\url{https://en.wikipedia.org/wiki/Power_iteration}},\ \bibinfo {note}
  {accessed: 2020-06-14}\BibitemShut {NoStop}%
\end{thebibliography}%

\end{document}